\documentclass[a4paper,12pt]{article}

\RequirePackage[OT1]{fontenc}
\usepackage{amsmath,natbib,amsfonts,amssymb,slashbox,multirow,rotating,array}
\usepackage{hhline}
\RequirePackage[colorlinks,citecolor=blue,urlcolor=blue]{hyperref}
\RequirePackage{hypernat}
\usepackage[dvips]{color}
\usepackage{graphics}
\usepackage{graphicx}
\usepackage[small]{caption}
\usepackage{titlesec}
\usepackage{setspace}
\usepackage{geometry}
\titleformat{\section}[block]{\large\bf\filcenter}{\thesection.}{1em}{}
\def\bs{\bigskip}
\def\cl{\centerline}
\def\ms{\medskip}
\def\ni{\noindent}

\def\A{{\rm A}}
\def\ab{\allowbreak}
\def\bX{{\bar X}}
\def\bY{{\bar Y}}

\def\cJ{{\cal J}}
\def\cS{{\cal S}}
\def\cT{{\cal T}}
\def\ep{\epsilon}
\def\ga{\gamma}
\def\hg{{\hat g}}
\def\hga{{\hat\ga}}
\def\hj{{\hat\jmath}}
\def\hSi{{\widehat\Si}}
\def\IR{{\rm I\!R}}
\def\mhf{^{-1/2}}
\def\mo{^{-1}}
\def\mi{\,|\,}
\def\ra{\to}
\def\rai{\ra\infty}
\def\si{\sigma}
\def\Si{\Sigma}
\def\sumi{\sum_i\,}

\def\T{^{{\rm T}}}
\def\thf{{\textstyle{1\over2}}}
\def\ds{\displaystyle}

\begin{document}

\begin{center}
{\LARGE\bf An Algorithm for Nonlinear, Nonparametric Model Choice and Prediction}

~\\~\\ {\large Fr\'ed\'eric FERRATY\footnote{\noindent Fr\'ed\'eric Ferraty is Professor, Toulouse Mathematics Insitute, University of Toulouse, 31062 Toulouse, France (E-mail: frederic.ferraty@math.univ-toulouse.fr). Peter Hall is Professor, Department of Mathematics and Statistics, University of Melbourne, Parkville 3010, Australia (E-mail: halpstat@ms.unimelb.edu.au)}, and Peter HALL}

\end{center}
~\\ 
\hspace*{1cm}\begin{minipage}[c]{0.85\textwidth} 
We introduce an algorithm which, in the context of nonlinear regression on vector-valued explanatory variables, chooses those combinations of vector components that provide best prediction. The algorithm devotes particular attention to components that might be of relatively little predictive value by themselves, and so might be ignored by more conventional methodology for model choice, but which, in combination with other difficult-to-find components, can be particularly beneficial for prediction. Additionally the algorithm avoids choosing vector components that become redundant once appropriate combinations of other, more relevant components are selected.  It is suitable for very high dimensional problems, where it keeps computational labour in check by using a novel sequential argument, and also for more conventional prediction problems, where dimension is relatively low. We explore properties of the algorithm using both theoretical and numerical arguments.
\end{minipage}
~\\
\hspace*{1cm}\begin{minipage}[c]{0.85\textwidth}  ~ 
\end{minipage}

\section{INTRODUCTION}

\ni For more than 30 years statisticians have sought to identify the relevant vector components in relatively high-dimensional prediction problems.  Today, in the case of data from fields such as genomics, astronomy and consumer preference modeling, the challenges are greater than in the past, with the ratio of dimension to sample size often being higher than every before.  In the present paper we suggest a new, highly adaptive algorithm that can be used to build predictive models in both contemporary and classical settings.  Our approach is designed specifically for cases where the response is a nonlinear function of the predictors, and where we wish to be economical in our choice of variables.  

Particularly in cases where dimension is greater than sample size, a great deal of attention has been devoted in the last 15 years to model choice in the framework of linear models. In this setting, Tibshirani's (1996) lasso was the starting point for the development of many techniques: coordinate descent methods (Fu, 1998, Friedman {\em et al.}, 2007), smoothly clipped absolute deviation (Fan and Li, 2001), least angle regression (Efron {\em et al.}, 2004), elastic net (Zou and Hastie, 2005), adaptative lasso (Zou, 2006), Dantzig selector (Cand\`es and Tao, 2007), relaxed lasso (Meinshausen, 2007), group lasso (Yuan and Lin, 2008), multi-step adaptative lasso (B\"uhlmann and Meier, 2008). Overviews of this work have been provided by Hastie {\em et al.}~(2009), Fan and Lv (2010) and B\"ulhmann and van de Geer (2011). 

These variable selection tools have been applied successfully to various high-dimensional datasets, but their effectiveness can be hindered by the assumption of a linear relationship between response and covariates. One problem is that the high-dimensional setting makes it difficult to validate the existence of the linear relationship. Moreover, it is common to encounter nonlinear structure even in standard, relatively low-dimensional multivariate regression models, and there is no a priori reason why such structure should not occur in high-dimensional cases. 

However, it can be very challenging to investigate nonlinear relationships when there are many variables.  There exists a literature on additive modeling, which often is treated as an extension of the lasso by combining the group lasso with basis expansion of each one-dimensional additive component.  See, for example, the work of Meier {\em et al.}~(2009), Ravikumar {\em et al.}~(2009) and Huang {\em et al.}~(2010).

Ferraty {\em et al.}~(2010) endeavoured to go beyond these techniques by developing methodology that captures interactions, using a stepwise forward search algorithm founded on minimizing a cross-validation criterion.  However, although this approach enjoys good performance in many cases, it fails in a worrying number of cases, where small submodels are not detected.   

The new algorithm suggested in this paper is based on enlarging the class of possible combinations of covariates retained at each step, while keeping the run time within reasonable bounds.  This is an important issue from a pratical viewpoint. Our methodology is given in Section~2, where our approach to building and selecting submodels is discussed first in overview and then described in detail.  The technique is illustrated in Section~3 by application to a real genomics dataset, and in Section~4 in a simulation study.  Theoretical issues are treated in Section~5.	  
\bs

\section{METHODOLOGY}

\subsection{Measuring mean squared variation.} 

Given independent and identically distributed data pairs $(X_i,Y_i)$ for $i\in\cS=\{1,\ldots,n\}$, where $X_i=(X_{i1},\ldots,X_{ip})$ is a $p$-vector and $Y_i$ is a scalar, we wish to choose a small number of vector components, or variables or features, of $X_i$ on which to regress $Y_i$, with the aim of predicting a future $Y$ for a given~$x=(x_1,\ldots,x_p)$. 

Our methodology is built around an algorithm, discussed in Section~2.2 and defined concisely in Section~2.3, for determining the extent to which a given subsequence,  $X_{ij_1},\ldots,X_{ij_\ell}$ say, of the components of $X_i$ successfully predict~$Y_i$.  Each step of the algorithm involves using our favorite nonparametric function estimator, for example a local linear approach or a spline, to construct a predictor $\hga_{j_1,\dots,j_\ell}(x_{j_1},\ldots,x_{j_\ell})$ of $Y$ from the dataset $\{(X_{ij_1},\ldots, X_{ij_\ell},Y_i);~i\in\cS\}$, where $(x_{j_1},\ldots,x_{j_\ell})$ is a subvector of $x$.  Then compute the standard cross-validation criterion 
$$
S(j_1,\ldots,j_\ell)=\sum_{i=1}^n \: \{Y_i-\hga_{j_1,\ldots,j_\ell}^{-i}(X_{ij_1},\ldots,X_{ij_\ell})\}^2\,
w_\ell(X_{ij_1},\ldots,X_{ij_\ell})\,,\eqno(2.1)
$$
which measures the success of $\hga_{j_1,\ldots,j_\ell}^{-i}(X_{ij_1},\ldots,X_{ij_\ell})$ in predicting~$Y_i$ where $\hga_{j_1,\ldots,j_\ell}^{-i}$ is the leave-one-out estimator derived from $\cS\mbox{\textbackslash}\{i\}$.  The function $w_\ell$ in (2.1) is taken to be nonnegative.  

In order to simplify notation, let $\cJ=\{j_1,\ldots,j_\ell\}$ be a subset of  $\{1,\ldots,p\}$ so that, for any $p$-dimensional vector $u=(u_1,\ldots,u_p)$ of $\mathbb{R}^p$, $u^\cJ$ stands for the subvector $(u_{j_1},\ldots,u_{j_\ell})$.   Then, (2.1) may be written in an equivalent way as
$$
S(\cJ)=\sum_{i\in\cT} \: \{Y_i-\hga_{\cJ}^{-i}(X_i^{\cJ})\}^2\,
w_{|\cJ|}(X_i^{\cJ})\,,\eqno(2.2)
$$
where $|\cJ|$ is the size of $\cJ$.

If $\cJ_1,\ldots,\cJ_k$ are distinct subsets of indices then the permutation of $\cJ_1,\ldots,\cJ_k$ that is used in each of the steps in Section~2.3, for different values of $k$, is that which places the values of $S(\cJ_1),\ldots,S(\cJ_k)$ in increasing order.  In the subsequent step of the algorithm we merge $\cJ_1,\ldots,\cJ_k$ in a pairwise manner, creating new subsets of indices $\cJ_1\cup\cJ_2$, $\ldots$, $\cJ_1\cup\cJ_k$, $\ldots$, $\cJ_{k-1}\cup\cJ_k$ that are rearranged again to rank the corresponding predictive values; and we repeat this process until we obtain a subset $\cJ$ with a sufficiently small value of~$S(\cJ)$.  

\subsection{Overview of algorithm.}

The first step of the algorithm involves searching over all single subsets $\{1\},\ldots,\{p\}$, the next over all combinations $\{j,j'\}$, the third over all  combinations of the previous ones (i.e.\ $\{j_1, j_1'\}\cup\{j_2, j_2'\}$), and so on. Normally this would be prohibitively expensive from a computational viewpoint.  Indeed, in many problems doing even the $O(p^2)$ search over pairs of indices would be out of the question.  However, we use the following ``trick'' to reduce labour.  Having searched over single subsets and ranked the variables there, we look only at the top $\sqrt{p}$ variables when constructing the sets $\{j,j'\}$ over which we search in the next step.  There are only $O(\sqrt{p}^{2})=O(p)$ subsets of indices constructed in this way, and so the search over sets $\{j, j'\}$ is not much more onerous than it was in the case of the single subsets.  

In Section~2.3 we note that $O(p)$ may not, in general, be a good description of the upper bound to the capability of our computational resources.  Instead we take $O(q)$ to be that bound, where $q$ might be larger than $p$ if our resources are relatively extensive, or less than $p$ if the inherent multiplier of a power of $n$, which for simplicity we omitted from the arguments above, is problematic.  In this case our algorithm ``sniffs out'' the trace of potentially significant variables among the first $\sqrt{q}$ variables when building bivariate predictors, and subsequently also when constructing predictors of higher order.  For now, however, we assume that $q=p$.

It should be stressed that the steps in our algorithm rely on the variables that are ``useful'' for prediction making themselves known, to at least some extent, when we are experimenting with prediction based on a single variable.  Experimentation is described in Step~1 in Section~2.3.  Variables that are useful for building higher-order predictors do not have to be present in the top few of the $p$ variables, but some of them should be apparent with sufficient strength to lie among the top $\sqrt{p}$ variables.  It is difficult to see how this constraint can be removed without using a relatively a crude, model-based approached to variable selection.  The advantage of our alternative approach is that, if a variable shows itself to be just slightly useful for prediction in isolation, in particular if it lies among the top $\sqrt{p}$ variables, then we have an opportunity to detect its importance even if its main contributions are felt only when it operates in conjunction with 
 one or more other variables.  In contrast, conventional approaches to feature selection, based on linear models, can completely overlook variables that have a major impact only through interaction with one or more other variables.  

\subsection{Details of algorithm.}

\ni{\sl Step~1: Prediction based on a single variable.}~Consider the $p$ singletons $\cJ_1=\{1\},\ldots,\cJ_p=\{p\}$, and compute the permutation $\hj_1(1),\ab\ldots,\hj_1(p)$ of the indices $1,\ldots,p$ that represents the ranking $S\{\cJ^1(1)\}\leq\ldots\leq S\{\cJ^1(p)\}$, with $\cJ^1(k)=\cJ_{\hj_1(k)}$ for $1\leq k\leq p$ and where $S$ is defined as at~(2.1).  If $\hj_1(k_1)<\hj_1(k_2)$ then $X_{i\hj_1(k_1)}$ better explains $Y_i$, in a particular sense, than does $X_{i\hj_1(k_2)}$.  In this sense, a regression of $Y$ on the $\hj_1(1)$th component of $X$ produces the ``best'' predictor based on a single variable.  

\ni{\sl Step~2: Prediction based on two variables.} Assume that our computing resources are limited to $O(q)$ calculations, multiplied by a low power of $n$, and put $p_1=\sqrt{q}$. From the top $p_1$ subsets $\cJ^1(1),\ldots,\cJ^1(p_1)$, build the set of all $p_2^*=\thf\,p_1\,(p_1-1)=O(q)$ pairs $\cJ_1=\cJ^1(1)\cup\cJ^1(2),\ldots,\cJ_{p_1-1}=\cJ^1(1)\cup\cJ^1(p_1),\cJ_{p_1}=\cJ^1(2)\cup\cJ^1(3),\ldots,\cJ_{p_2^*}=\cJ^1(p_1-1)\cup\cJ^1(p_1)$. Then, compute the permutation $\hj_2(1),\ab\ldots,\hj_2(p_2^*)$ of the indices $1,\ldots,p_2^*$ that places the values $S(\cJ_{\hj_2(k)})$, for $1\leq k\leq p_2^*$, in increasing order, and retain for the next step only the $p_2=p_1=\sqrt{q}$ top subsets $\cJ^2(1)=\cJ_{\hj_2(1)},\ldots,\cJ^2(p_2)=\cJ_{\hj_2(p_2)}$.  A regression of $Y$ on $X^{\cJ^2(1)}$ provides the ``best'' predictor based on just two variables.  

\ni{\sl Steps~3,4,$\ldots$: Prediction based on $\ell\geq 3$ variables.} In step 1, or respectively step 2, the procedure builds only singletons, or respectively pairs. However, in step $\ell\geq 3$ the algorithm may generate subsets $\cJ$ of indices such that $\ell\leq|\cJ|\leq 2^{\ell-1}$. For instance, if we consider the sets $\cJ^2(1)=\{j_1,j_2\},~\cJ^2(2)=\{j_1,j_3\},~\cJ^2(3)=\{j_2,j_4\},~\ldots$, the third step of our algorithm will build a new family of subsets containing $\cJ^2(1)\cup\cJ^2(2)=\{j_1,j_2,j_3\},\cJ^2(1)\cup\cJ^2(3)=\{j_1,j_2,j_4\},\ldots,\cJ^2(2)\cup\cJ^2(3)=\{j_1,j_2,j_3,j_4\},\ldots$, which produces subsets of size 3 or 4.  Assume we have constructed, in the previous step, an ordered sequence of subsets $\cJ^{\ell-1}(1),\ldots,\cJ^{\ell-1}(p_{\ell-1})$ where all indices of each subset are listed in increasing numerical order and the subsets are ordered so that the corresponding values of $S\{\cJ^{\ell-1}(j)\}$ are increasing. The new family of subsets
  
 $\cJ_1=\cJ^{\ell-1}(1)\cup\cJ^{\ell-1}(2),\ldots,\cJ_{p_{\ell-1}-1}=\cJ^{\ell-1}(1)\cup\cJ^{\ell-1}(p_{\ell-1}),~\cJ_{p_{\ell-1}}=\cJ^{\ell-1}(2)\cup\cJ^{\ell-1}(3),\ldots$ is filtered in order to retain only $p_\ell^*$ distinct subsets where $p_\ell^*\leq \thf\,p_{\ell-1}\,(p_{\ell-1}-1)$, and the indices in each subset form a strictly increasing sequence. Then, the permutation $\hj_\ell(1),\ldots,\hj_\ell(p_\ell^*)$ of $1,\ldots,p_\ell^*$ is carried out so that $S(\cJ_{\hj_\ell(1)})\leq\ldots\leq S(\cJ_{\hj_\ell(p_\ell^*)})$, and we retain for the next step only the $p_\ell=\min(p_1,p_\ell^*)$ top subsets $\cJ^\ell(1)=\cJ_{\hj_\ell(1)},\ldots,\cJ^\ell(p_\ell)=\cJ_{\hj_\ell(p_\ell)}$.  

The algorithm can be terminated when a predetermined percentage of the mean squared variation among the $Y_i$s is explained by the regressions, or when the difference between two successive measures of that variation falls below a given level, or there is a marked ``kink'' in a graph of the minimum value of $S\{\cJ^\ell(1)\}$ against~$\ell$.  The second of these three rules can be interpreted as stopping as soon as, for some $\ell\geq1$,
$$
\frac{S\{\cJ^\ell(1)\}-S\{\cJ^{\ell+1}(1)\}}{S\{\cJ^\ell(1)\}}\leq t\,,\eqno(2.3)
$$
where $t=t(n)$ is a user-choosable threshold expressing a necessary minimum gain in going to the next step.  The estimator $\hg$ is then computed in a standard way, using the ``favorite nonparametric function estimator'' referred to in Section~2.1, from the data pairs $(X_i^{\cJ^\ell(1)},Y_i)$ for $1\leq i\leq n$, where $X_i^{\cJ^\ell(1)}=(X_{ij})_{j\in\cJ^\ell(1)}$.  The ``kink'' approach is commonly used to determine a stopping point for clustering algorithms, where the value of $S$ at (2.1) is replaced by a measure of the tightness of a cluster.  

 From now on, this nonparametric variable selection method will be referred to as NOVAS. 

\subsection{Practical issues.}

\begin{figure}[h]
\begin{center}
\includegraphics[height=6cm, width=8cm]{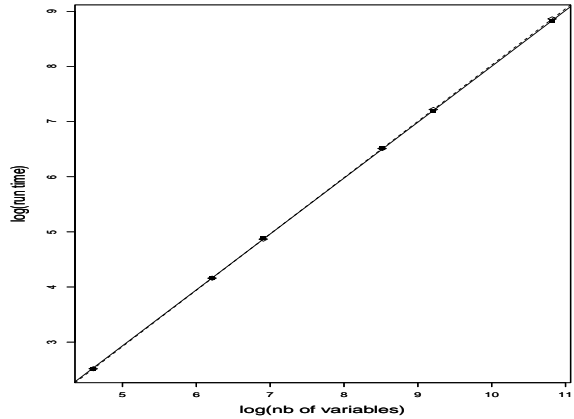}
\caption{Linear fits based six points, depicted by the solid line, and only the first two points, indicated by the dashed line.}
\label{fig1}
\end{center}
\end{figure}

Our method is computationally intensive; launching it with a very large dataset may be time consuming. One way to speed up computation is to parallelize the algorithm. Indeed, as soon as a computer is equipped with a multicore processor, which is the case for most current computers, parallelization allows us to process independent tasks simultaneously. The running time is then divided by the number of independent tasks that the multicore processor is able to manage. The programming language R (R Development Core Team, 2011) offers packages that make such a parallelization easy; see for instance the R package ``doSNOW'' of Revolution Analytic (2011). In addition, since R is freeware and used intensively by academic researchers, this programming language is one of the most popular in the statistical community. For these reasons we decided to use the R programming language to implement our variable selection method. All results presented with respect to the real dataset application (see Section \ref{sec:genomicsdata}) were obtained using a laptop with a 4-core, 2 GHz processor with 4 Go RAM. To give an idea about the run time, the R routine NOVAS is repeated for an artificial dataset containing $p=$100, 500, 1,000, 5,000, 10,000 and 50,000 covariates in such a way that $\ell=4$ steps are run systematically for each $p$ with $n=100$ and default threshold parameter $t=0.05$. Seven parallel jobs are launched (which is the more efficient number of parallel jobs for the used laptop); the corresponding run times (in seconds) are displayed in Figure \ref{fig1}; for instance, NOVAS lasts 691s when $p=10000$. It is worth noting an almost perfect linear relationship between the log number of variables, i.e.\ values of $p$, and the corresponding log run time. 
As suggested there, considering only $p=100$ and $p=500$ is enough to gain a good approximation to the run time for much higher dimensional cases. The simulation study, which requires massive computations, was granted access to a supercomputer resource (see the acknowledgements). 

The nonparametric regression estimator $\hga_{\cJ}$,  introduced in (2.1), is the usual local linear one (see e.g.\ Fan and Gijbels, 1996). In order to speed up the computations, the covariates are standardized and for each subset $\cJ$, we chose a common bandwidth among given a set of bandwidths to minimize the cross-validation criterion $S(\cJ)$ defined at (2.2) with $w_{|\cJ|}\equiv 1$.

\section{GENOMICS DATASET}\label{sec:genomicsdata}

This dataset was discussed by Bushel {\em et al.} (2007).  It is also addressed in the R package mixOmics, which is designed to explore and integrate omics data and was developed by Dejean {\em et al.}~(2011). The dataset treats liver toxicity and contains the expression levels of 3116 genes, or covariates, and nine clinical measurements, or scalar responses, for 64 rats. The original dataset included supplementary clinical responses, but these were not used since only three distinct values were available. 

\begin{figure}[h]
\centering
\makebox{\includegraphics[height=12cm,width=14cm]{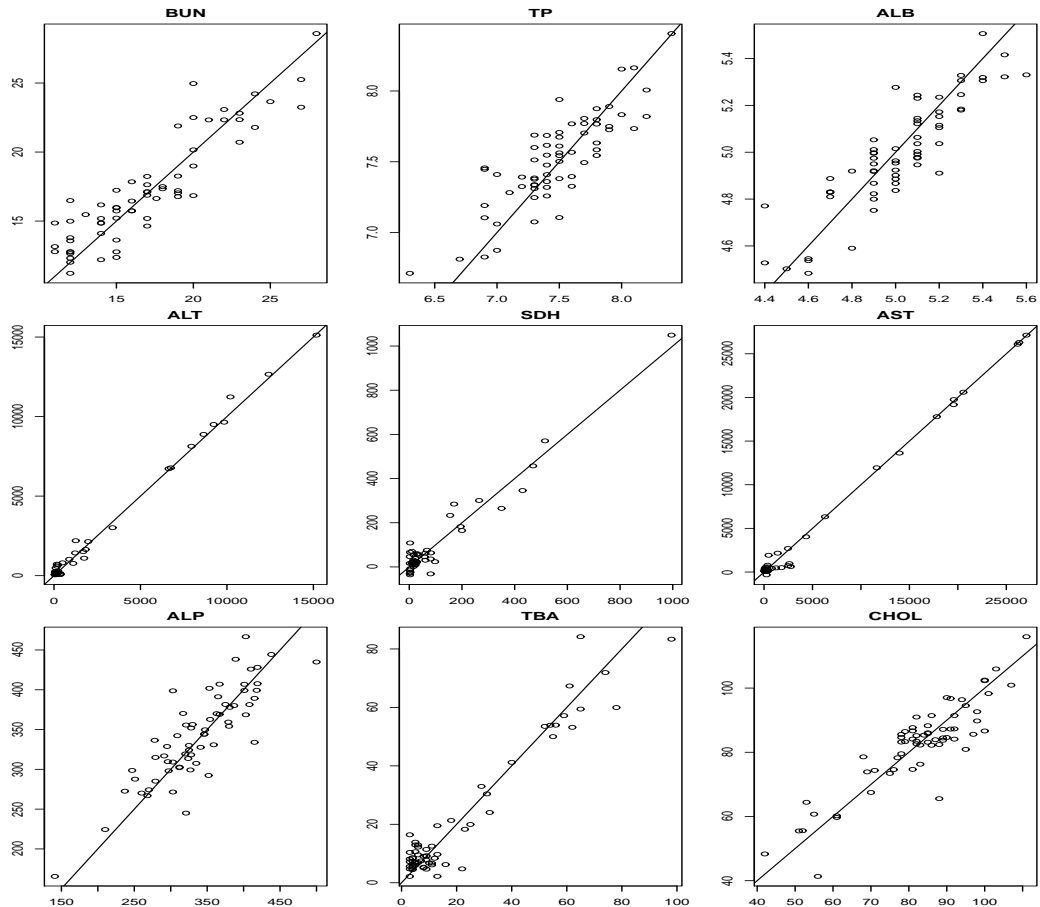}}
\caption{\label{fig2}Observations (horizontal axis) plotted against leave-one-out predictions for each clinical measurement.}
\end{figure}

Our aim was to select, for each scalar response, the genes leading to the best predictor in terms of the cross-validation criterion at~(2.1). Table \ref{tab1} details stages of NOVAS when the aim is to predict the level of urea nitrogen. As pointed out earlier, the final model may involve variables not necessarily identified as the most predictive ones in the previous stages.  
\begin{table}[h]
\caption{\label{tab1}The table gives, at each stage, the best predictive subset of variable number(s) and corresponding leave-one-out cross-validation criterion.}
\centering
\begin{tabular}{crc}
Stage number & Selected genes numbers & cv\\~\\ \hline
1 & 1165 & 6.83\\ \hline
2 & 1866\ 2050 & 5.22\\ \hline
3 & 1000\ 1167\ 1837\ 1957 & 3.76\\ \hline
4 & 1000\ 1167\ 1837\ 1899\ 1957 & 3.27\\ \hline
\end{tabular}
\end{table}
Table \ref{tab2} gives, for each clinical measurement, the gene numbers, i.e.\ the subset $\widehat{\cJ}$, selected by NOVAS, together with values of the corresponding cross-validation criterion, i.e.\ $S(\widehat{\cJ})$, 
\begin{table}[h]
\caption{\label{tab2}NOVAS selected models for each clinical measurement.}
\centering
\begin{tabular}{c m{8cm}}
Clinical measurement & Selected genes numbers \\~\\ \hline
BUN & 1000\ 1167\ 1837\ 1899\ 1957 \\ \hline
TP & 1159\ 1970\ 2020\ 2173\ 2923\  2927\ 2971 \\ \hline
ALB & 1038\ 1165\ 1992\ 2020\ 2105\ 2669\ 2867\ 2921 \\ \hline
ALT & 1846\ 1871\ 1883\ 1909\ 1910\ 1911\ 1915\ 1921\ 2042 \\ \hline
SDH & 764\ 1145\ 1624\ 1866\ 1940\ 1992\ 1996\ 2894 \\ \hline
AST & 977\ 1116\ 1161\ 1335\ 1826\ 1891\ 1909\ 1961\ 2197\ 2201 \\ \hline
ALP & 1064\ 1484\ 1817\ 1823\ 2007\ 2385\ 2819 \\ \hline
TBA & 1891\ 1913\ 1916\ 1917\ 1954\ 2200\ 2205 \\ \hline
CHOL & 1836 1875 2044 \\ \hline
\end{tabular}
\end{table}
where the clinical response abbreviations were defined as follows: BUN, urea nitrogen;  TP, total protein; ALB, albumin; ALT, alanine aminotransferase; SDH, sorbitol dehydrogenase; AST, aspartate aminotransferase; ALP, alkaline phosphatase; TBA, total bile acids; and CHOL, cholesterol. Two clinical responses, BUN and CHOL, require three or five genes. Other clinical measurements, including AST, involve many more genes. Figure \ref{fig2} displays, for each clinical variable, the observed values plotted against the leave-one-out predictions. As can be seen, clinical responses are well explained by the selected genes. 

An important question arises: is the quality of the leave-one-out predictions high? To answer this question we propose comparing the leave-one-out cross-validation criterion, i.e.\ $S(\widehat{\cJ})$, obtained by NOVAS, with various alternative predictive methods: 
\begin{itemize}
\item[$\bullet$] partial least squares regression, PLS, which is a non-selective iterative linear method and derives successive linear combinations, or loadings, of covariates maximizing its correlation with the response. It was originally developed by Wold (1966) for applications in economics and became a popular tool in the chemometrics community; see, for instance, Geladi and Kowalski (1986) or Martens and Naes (1989);
\item[$\bullet$] a sparse version of PLS, sPLS, including a lasso step leading to sparse loadings, developed by L\^e Cao {\em et al.} (2008), who experimented with this genomics dataset;
\item[$\bullet$] least angle regression, LAR, introduced by Efron {\em et al.} (2004), which is one of the most popular selective linear regression methods;
\item[$\bullet$] most predictive design points (MPDP), which is also an existing nonparametric alternative method and which we shall discuss in Section~4.5.
\end{itemize}
The PLS method requires the choice of only one parameter, the total number of loadings, whereas sPLS needs several variables, specifically the total number of loadings and the sparsity expressed as the number of zeros for each loading. The LAR procedure requires choice of the optimal fraction of non-zero values in the vector of parameters. For all these competing methods, the parameters were optimized so as to minimize the predictive leave-one-out criterion, and Table 3 gives the smallest leave-one-out cross-validation values obtained for each procedure. 
\begin{table}
\caption{\label{tab3}Leave-one-out cross-validation values for each clinical measurement, or response, and each method.  Minimum values in each row are given in bold.}\centering
\begin{tabular}{cccccc}
\backslashbox{Responses}{Methods} & PLS & sPLS & LAR & MPDP & NOVAS \\ \hline 
BUN  & 6.62 & 8.106 &  8.67 & {\bf 2.62} & 3.27\\ \hline
TP  & 0.104 & 0.115 & 0.117 & 0.072 & {\bf 0.045}\\ \hline
ALB  & 0.0351 & 0.0378 & 0.044 & 0.02 & {\bf 0.015}\\ \hline
ALT  & 1236163 & 1260917 & 1709814 & {\bf 46834} & 60621\\ \hline
SDH & 19736.38 & 23512.95 & 21314.9  & 2669.6 & {\bf 1404.7}\\ \hline
AST  & 5232362 & 6131264 & 92534432 & 2580580 & {\bf 318682}\\ \hline
ALP  & 3097.81 & 3194.96 & 3203.1 & 1225.4 & {\bf 1043.7}\\ \hline
TBA & 138.26 & 118.63 & 153.12 & 67.50 & {\bf 39.73}\\ \hline
CHOL & 76.91 & 72.87 & 94.96 & {\bf 26.77} & 40.68\\ \hline
\end{tabular}
\end{table}
It can be seen that NOVAS outperforms alternative linear methods, since it is able to take nonlinearities into account.

\begin{figure}[h]
\centering
\makebox{\includegraphics[height=12cm, width=14cm]{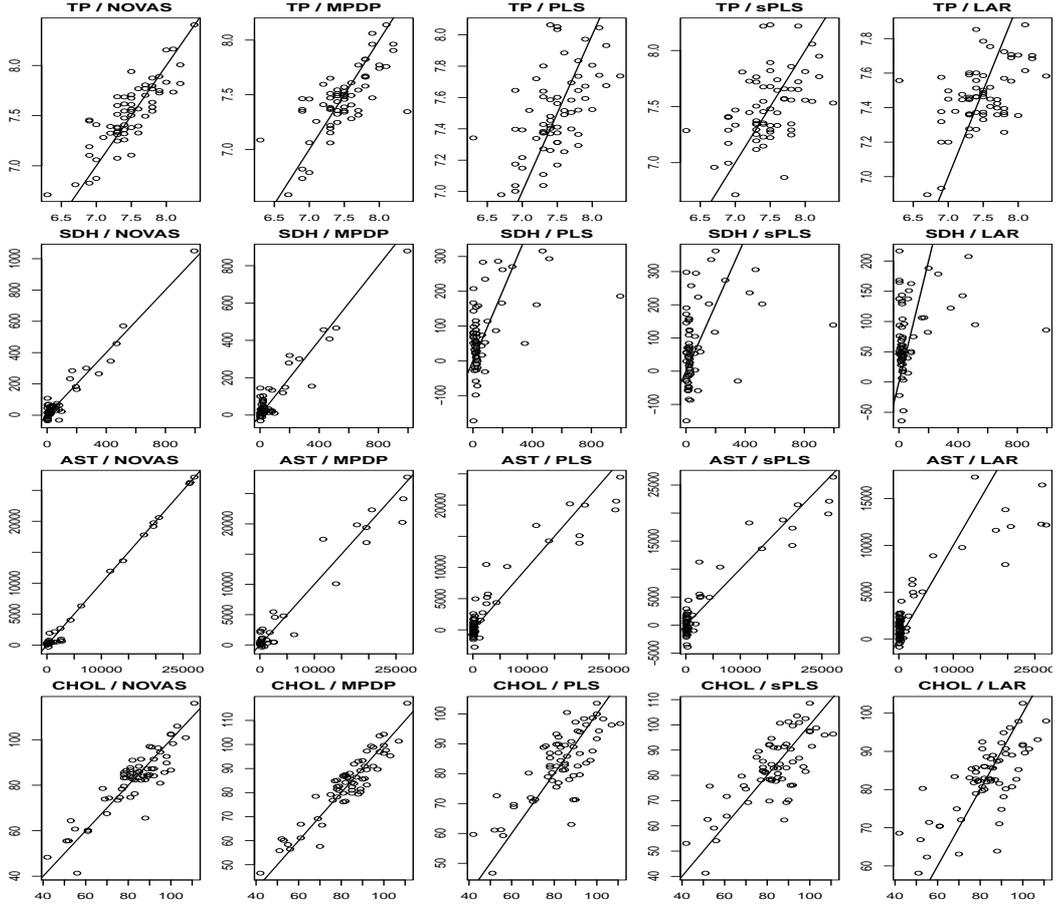}}
\caption{\label{fig3}Four observed responses, TP, SDH, AST and CHOL, against leave-one-out estimations.}
\end{figure}

To enable predictive performances to be visualized, Figure \ref{fig3} compares, for each method, the results of leave-one-out estimation applied to a sample of four clinical measurements: TP, SDH, AST and CHOL. Clearly, the two nonparametric selective procedures, NOVAS and MPDP, have significantly greater predictive performance than the linear procedures, and NOVAS is much the stronger of the two.  For example, NOVAS leads to more accurate predictions in most cases, and enjoys spectacular performance when applied to predicting the clinical measurement AST. 

\section{ASSESSING PERFORMANCE}

\subsection{Simulated regression models.}  

We consider five models, indexed by a superscript $m$ in square brackets and having the form
$$
Y_i \ =\ \gamma_{1,2,3}^{[m]} (X_{i1},X_{i2},X_{i3})\, +\, \varepsilon_i^{[m]},\ i=1,\ldots,n_m, 
$$
where 
$$
\begin{array}{ccl}
\gamma_{1,2,3}^{[1]}(X_{i1},X_{i2},X_{i3}) & = & X_{i1}^2+X_{i2}^2+X_{i3}^2\:, \\~\\
\gamma_{1,2,3}^{[2]}(X_{i1},X_{i2},X_{i3}) & = & \left|X_{i1}X_{i2}\right|+\left|X_{i1}X_{i3}\right| +\left|X_{i2}X_{i3}\right|,\\~\\
\gamma_{1,2,3}^{[3]}(X_{i1},X_{i2},X_{i3}) & = & \left|X_{i1}X_{i2}X_{i3}\right|, \\~\\
\gamma_{1,2,3}^{[4]}(X_{i1},X_{i2},X_{i3}) & = &\displaystyle \frac{\left|X_{i1}X_{i2}\right|+X_{i3}^2}{2+X_{i1}X_{i2}X_{i3}}\:,\\~\\
\gamma_{1,2,3}^{[5]}(X_{i1},X_{i2},X_{i3}) & = &\displaystyle \frac{\left|X_{i1}X_{i2}\right|+\left|X_{i1}X_{i3}\right|}{2+\left|X_{i2}X_{i3}\right|}\:.
\end{array}
$$
The vector components $X_{ij}$ are taken to be independent and identically distributed as uniform $[-1,1]$, and the errors $\varepsilon_i^{[m]}$ are independent and identically distributed as normal $N(0,\sigma_m^2)$, where $\sigma_m^2=0.05\,{\rm var}\{\gamma_{1,2,3}^{[m]}(X_{i1},X_{i2},X_{i3})\}$.  
Different sample sizes will be considered in the simulation study to take into account the varying complexities and dimensionalities of these models.

\subsection{Influence of the sample size, $n$, and the number, $p$, of covariates.}\label{sec:influence-n-p}  

Is our NOVAS procedure able to recognize the correct subset $\cJ=\{1,2,3\}$, even for large values of~$p$?  To answer this question our selected procedure was launched on each of the five models with four sample sizes $n=50,\,100,\,150,\,200$, three different sets of covariates of size $p=100,\,1000,\,10000$ and the threshold parameter $t$, defined at (2.3), set equal to 0.05 for all runs. We then consider $5\times4\times 3$ situations and the simulation scheme was repeated 100 times, producing 100 datasets for each situation. Table \ref{tab4} presents the results in order of sample size. One remarks that higher is the complexity of the model, larger has to be the sample size for recognizing it correctly. So, the role played by the  sample size $n$ corroborates what happens usually in statistics; higher is the dimensionality of the model, larger has to be the sample size for getting good estimation. For a too small sample size (i.e. $n=50$), NOVAS is not able to recognize models with a good frequency excepted Model 1 when $p=100$. In the opposite, when considering a large sample size (i.e. $n=200$), one gets good results for all models, even for large set of covariates. When focusing on the influence of the number $p$, its impact on the behaviour of NOVAS is clear: higher is the size of the set of covariates, lower is the frequency of selecting the correct model. However, as soon as one considers a sample size for each model large enough, the influence of $p$ is not so dramatic; for any $p=100,\,1000,\,10000$, one gets good and stable results for models 1 and 2 when $n=100$, model 3 when $n=150$, and models 4 and 5 when $n=200$. 
\begin{table}[h]
\caption{\label{tab4}Number of times, out of 100, that NOVAS selected the correct model.}
\centering
\begin{tabular}{lll}
\begin{tabular}{c} 
 \\
$p$  \\ \hline\hline 
Model 1 \\ \hline
Model 2 \\ \hline
Model 3 \\ \hline
Model 4 \\ \hline
Model 5\\ \hline
\end{tabular}
&\begin{tabular}{ccc} 
\multicolumn{3}{c}{$n=50$}\\
100 & 1000 & 10000 \\ \hline\hline 
84 & 46 & 12 \\ \hline
49 & 25 & 1 \\ \hline
4 & 2 & 0 \\ \hline
18 & 6 & 0 \\ \hline
2 & 1 & 0\\ \hline
\end{tabular}
& \begin{tabular}{ccc} 
 \multicolumn{3}{c}{$n=100$}\\
  100 & 1000 & 10000 \\ \hline\hline 
  99 & 100 & 100 \\ \hline
  100 & 99 & 97 \\ \hline
  89 & 79 & 56 \\ \hline
  78 & 58 & 34 \\ \hline
  46 & 28 & 1\\ \hline
\end{tabular}
\\ ~\\ 
\begin{tabular}{c} 
 \\
$p$  \\ \hline\hline 
Model 1 \\ \hline
Model 2 \\ \hline
Model 3 \\ \hline
Model 4 \\ \hline
Model 5\\ \hline
\end{tabular}
&\begin{tabular}{cccc} 
\multicolumn{3}{c}{$n=150$}\\
100 & 1000 & 10000 \\ \hline\hline 
100 & 100 & 100 \\ \hline
100 & 100 & 100 \\ \hline
100 & 100 & 99 \\ \hline
97 & 91 & 74 \\ \hline
76 & 67 & 54\\ \hline
\end{tabular}
& \begin{tabular}{ccc}
\multicolumn{3}{c}{$n=200$}\\
 100 & 1000 & 10000 \\ \hline\hline 
 100 & 100 & 100 \\ \hline
 100 & 100 & 100 \\ \hline
 100 & 100 & 100 \\ \hline
 100 & 98 & 95 \\ \hline
 97 & 90 & 76\\ \hline
\end{tabular}
\end{tabular}
\end{table}

\subsection{Influence of the threshold $t$.} 

According to the previous section, considering different sample sizes allows to reduce the effect of the dimensionality of the simulated models on NOVAS (see Table \ref{tab5}). A high value of $t$ tends to stop the procedure too early, in which case NOVAS would retain too small a set of variables. A low value of $t$ results in the selected model incorporating too large a number of variables. However, as we shall show, NOVAS is not particularly sensitive to the value of $t$. 
\begin{table}
\caption{\label{tab5}Number of times, out of 100, that NOVAS selected the correct model when $p=1000$.}
\centering
\begin{tabular}{cccccccccccc} 
$t$ & 0.01 & 0.05 & 0.1 & 0.15 & 0.2 & 0.25 & 0.3 & 0.35 & 0.4 & 0.45 & 0.5 \\ \hline\hline 
Model 1 ($n= 100$) &  100 & 100 & 100 & 100 &  100 & 98 & 95 & 72 & 46 & 12 & 8 \\ \hline
Model 2 ($n= 100$) &  99 & 100 & 99 & 100 &  98 & 87 & 69 & 41 & 25 & 5 & 1 \\ \hline
Model 3 ($n= 150$) &  100 & 100 & 99 & 99  & 94 & 88 & 61 & 44 & 9 & 1 & 0 \\ \hline
Model 4 ($n= 200$) &  98 & 98 & 99 & 94 & 69 & 38 & 12 & 6 & 1 & 0 & 0 \\ \hline
Model 5 ($n= 200$) &  92 & 90 & 87 & 89 & 84  & 75 & 45 & 17 & 2 & 1 & 0\\ \hline
\end{tabular}
\end{table}
As indicated in Table \ref{tab5}, there is a range of values for $t$, i.e.\ $t\leq0.2$, where NOVAS provides stable results. This encouraged us to use the default value $t=0.05$. 

\subsection{Influence of noise-to-signal ratio.} 

Noise-to-signal ratio is defined by ${\rm nsr}=\sigma_m^2/\mbox{var}\{\gamma_{1,2,3}^{[m]}\left(X_{i1},X_{i2},X_{i3}\right)\}$. Up to now, ${\rm nsr}=0.05$ has been used in our numerical experiments.  Table~\ref{tab6} summarises the influence of noise-to-signal ratio on the behaviour of NOVAS, and it can be seen that performance decreases by 10~to~54\% as noise-to-signal ratio increases by 100~to~700\%.  
\begin{table}[h]
\caption{\label{tab6}Number of times, out of 100, that NOVAS selected the correct model when $p=1000$.}
\centering
\begin{tabular}{ccccc} 
nsr & 0.05 & 0.1 & 0.2 & 0.4 \\ \hline\hline 
Model 1 ($n=100$)  & 100 & 100 & 100 & 78 \\ \hline
Model 2 ($n=100$) & 100 & 98 & 91 & 58\\ \hline
Model 3 ($n=150$) & 100 & 100 & 96 & 84\\ \hline
Model 4 ($n=200$) & 98 & 96 & 99 & 89\\ \hline
Model 5 ($n=200$) & 90 & 96 & 75 & 41\\ \hline
\end{tabular}
\end{table}
Nevertheless, the performance of NOVAS remains stable with respect to noise-to-signal ratio;  when ${\rm nsr}=0.1$, the ability of NOVAS to recognise the true subset is very good for all models, when ${\rm nsr}=0.2$, NOVAS is still largely correct for models 1 to 4, and when ${\rm nsr}=0.4$ the results for models 1, 3 and 4 are reasonable. The performance of NOVAS, and of the competing methods that we shall discuss in the next section, degrades further for higher values of noise-to-signal ratio.  

\subsection{Comparison with other methods.}

In this section, we compare NOVAS with the other competing methods introduced in Section~3: PLS, sPLS, LAR and MPDP which is a nonparametric selective technique  called ``most predictive design points,'' or MPDP proposed by Ferraty {\em et al.} (2010). Originally developed for functional data, the method remains valid in the more conventional high-dimensional setting of the present paper. The idea is to select, one by one, several variables among a large number of candidates in order to predict nonparametrically a scalar response. The first step of MPDP chooses the most predictive variable minimizing (2.1), and updates the subset of candidates by dropping it; the second step selects the most predictive variable among the new subset of candidates with respect to (2.1), and again updates the subset of candidates; and so on. This procedure is repeated until the relative gain in terms of the cross-validation criterion between two consecutive steps does not exceed some threshold; see (2.3). The nonparametric regression estimator suggested for MPDP is the local linear one. The fundamental difference with NOVAS comes at the second step; for any $\ell>1$, NOVAS may drop at step $\ell+1$ some covariates selected at step $\ell$ whereas it is not possible with the sequential feature of MPDP. 

In order to achieve this comparison study, a family of regression models $$Y_i\,=\,\gamma_{1,2,3}^{\alpha} (X_{i1},X_{i2},X_{i3})\, +\, \varepsilon_i$$ indexed with a scalar $\alpha$ is simulated with 
$$
\gamma_{1,2,3}^{\alpha} (X_{i1},X_{i2},X_{i3})\,=\,3\,+\,\alpha\,\left(X_{i1}+X_{i2}+X_{i3}\right)\,+\,(1-\alpha)\left(X_{i1}^2+X_{i2}^2+X_{i3}^2\right),
$$
where the $X_{ij}$'s and the $\varepsilon_i$'s are built according to the same scheme described in Section 4.1. We set the sample size $n=50$, the number of covariates $p=1000$, the noise-to-signal $snr=0.1$ and the threshold $t=0.05$. This family of models allow to consider pure nonlinear situation (i.e. $\alpha=0$) as well as pure linear setting (i.e. $\alpha=1$). We also take into account an intermediate semilinear models by setting $\alpha=0.35$; this value allows us to balance the variability due to the linear and nonlinear parts. Only variables 1, 2, and 3  are active. For each value of $\alpha$, 100 datasets are simulated. This particular simulation scheme tests severely the selective procedures since most of the time, they are not able to detect the exact set of active covariates (see  Table \ref{tab7}). LAR outperforms clearly both MPDP and NOVAS in the linear situation. However, in the pure nonlinear setting, the behavior of MPDP and NOVAS is much better than LAR (which was expected since LAR is not designed for nonlinear relationship) with a significant  advantage for NOVAS. 
\begin{table}[h]
\caption{\label{tab7}Number of times, out of 100, that the correct model is selected.}
\centering
\begin{tabular}{cccc} 
 & $\alpha=0$ (nonlinear) & $\alpha=0.35$ (semilinear) & $\alpha=1$ (linear)\\ \hline 
LAR & 0 & 2 & 50 \\ \hline
MPDP & 28 & 23 & 0\\ \hline
NOVAS & 44 & 51 & 0\\ \hline
\end{tabular}
\end{table}
In fact, most of the time, extra covariates outside the active ones are selected. In order to better assess the selective performance of these methods, Table \ref{tab8} details how many times, out of 100, each active covariate are selected, and this for each value of $\alpha$. Of course, when tabulating all selected models, in addition of active variables 1, 2, and 3, a quite large number of extra covariates are retained three times at most (see column "others" in Table \ref{tab8}).
\begin{table}[h]
\caption{\label{tab8}Number of times, out of 100, that indicated covariates are  selected; the column "others" gives the maximum number of times that a same extra covariate is selected over the 100 runs.}
\centering
\begin{tabular}{llll}
\begin{tabular}{c} 
 \\
Covariates \\ \hline\hline 
LAR \\ \hline
MPDP \\ \hline
NOVAS \\ \hline
\end{tabular}
&\begin{tabular}{cccc} 
\multicolumn{4}{c}{$\alpha=0$ (nonlinear)}\\
1 & 2 & 3 & others\\ \hline\hline 
0 & 1 & 0 & 3\\ \hline
48 & 51 & 52 & 2\\ \hline
70 & 69 & 70 & 2\\ \hline
\end{tabular}
& \begin{tabular}{ccccc} 
 \multicolumn{4}{c}{$\alpha=0.35$ (semilinear)}\\
  1 & 2 & 3 & others\\ \hline\hline 
  38 & 46 & 39 & 3 \\ \hline
  69 & 74 & 72 & 2\\ \hline
  82 & 84 & 89 & 2\\ \hline
\end{tabular}
& \begin{tabular}{ccccc} 
 \multicolumn{4}{c}{$\alpha=1$ (linear)}\\
  1 & 2 & 3 & others\\ \hline\hline 
  100 & 100 & 100 & 2 \\ \hline
  79 & 76 & 77 & 4\\ \hline
  93 & 90 & 93 & 3\\ \hline
\end{tabular}
\end{tabular}
\end{table}
Firstly, when focusing on the number of times that active covariates are detected, LAR outperforms NOVAS and MPDP in the linear case whereas NOVAS and MPDP works better in the semilinear and nonlinear setting. Secondly, for each value of $\alpha$, NOVAS better recognizes the active covariates than MPDP does. Thirdly, even in the linear setting, the ability of selecting correctly the active covariates for NOVAS is still high. 

Table~\ref{tab9} compares the predictive performance of NOVAS with all competing methods. To obtain these results the mean of the cross-validation estimator of average prediction error, defined at (2.1) and (2.2), was computed (over 100 simulated datasets for each value of $\alpha$).  It can be seen from the Table \ref{tab9} that PLS, sPLS and LAR perform similarly (although the predictive performance of PLS in the linear setting is of poor quality). Naturally, in the nonfavourable situation (i.e. nonlinear model), the linear methods fails; PLS, sPLS and LAR perform almost exactly the same as the simple leave-one-out empirical mean of the responses, where for any $i$, $Y_i$ is predicted naively by $(n-1)^{-1}\,\sum_{j\neq i}\,Y_j$. 
\begin{table}[h]
\caption{\label{tab9}mean and variance, in parentheses, of cross-validation criterion, out of 100 simulated datasets.}
\centering
\begin{tabular}{llll} 
 & $\alpha=0$ (nonlinear) & $\alpha=0.5$ (semilinear) & $\alpha=1$ (linear)\\ \hline 
PLS & 0.30 (0.004) & 0.27 (0.004) & 1.07 (0.037)\\ \hline
sPLS & 0.38 (0.023) & 0.25 (0.007) & 0.18 (0.004)\\ \hline
LAR & 0.28 (0.004) & 0.23 (0.003) & 0.21 (0.006)\\ \hline
MPDP & 0.11 (0.001) & 0.07 (0.001) & 0.15 (0.043)\\ \hline
NOVAS & 0.10 (0.0005) & 0.06 (0.0002) & 0.11 (0.004)\\ \hline
\end{tabular}
\end{table}
When comparing NOVAS with MPDP, it seems that both methods have similar predictive behavior with an advantage for NOVAS which induces smaller variances in all cases. 

But the superiority that NOVAS enjoys relative to MPDP include an ability to correctly identify the variables on which the regression actually depends with a shorter running time. Indeed, it is easy to see that, for achieving $k$ steps when dealing with a set of $p$ covariates, MPDP needs the estimation of $kp-0.5k(k-1)+1$ regression models whereas NOVAS involves only $kp/2$ ones (i.e. twice less). Here, for this comparative study, MPDP requires on average one step more than NOVAS which implies that the overall MPDP run time is at least two times longer than the NOVAS one. Another difference between MPDP and NOVAS appears in the situation when one has to deal with redundant variables that were correlated with non-redundant ones. To illustrate this aspect, we simulated datasets from model 4 with different sample sizes as in Section \ref{sec:influence-n-p} ($n=50,100,150,200$), noise-to-signal sets to 0.05, with a number of covariates equals to 1000. We replaced the 1000th explanatory variable by a combination of both the first two:  for $i=1,\ldots,n$, $X_{i\,1000}=X_{i\,1}^2|\,X_{i\, 2}|^{1/3}$. This 1000th variable contains redundant information which can mask the main role playing by variables 1 and 2 in the simulated regression model. Table \ref{tab10} details the selected variables, over 100 runs for each sample size. 
{\small
\begin{table}[h]
\caption{\label{tab10}Number of times, out of 100, that the indicated subsets were selected; $\cJ_{no\,intruder}$ is $\{1,3,1000\}$ either $\{2,3,1000\}$ or $\{3,1000\}$ and $\cJ_{intruder}$ represents any subset containing intruder(s) (i.e. $j$ with $j\notin\{1,2,3, 1000\}$).}\centering
\begin{tabular}{ccccccccc}
 & \multicolumn{2}{c}{$\{1,2,3\}$} &  \multicolumn{2}{c}{$\{1,2,3,1000\}$} & \multicolumn{2}{c}{$\cJ_{no\,intruder}$} & \multicolumn{2}{c}{$\cJ_{intruder}$}\\ \hline
n & MPDP & NOVAS & MPDP & NOVAS & MPDP & NOVAS & MPDP & NOVAS\\ \hline
50 & 1 & 1 & 0 & 0 & 6 & 19 & 93 & 80\\ \hline
100 & 7  & 38  & 2  & 1  & 17  & 33  & 74  & 28\\ \hline
150 & 2  & 79  & 18  & 2  & 19  & 15  & 71  & 4\\ \hline
200 & 0  & 96  & 43  & 4  & 23  & 0  & 34  & 0\\ \hline
\end{tabular}
\end{table}
}
Note that the ability of MPDP to identify variables 1, 2, and 3, but most of the time combined with variable 1000, increases with $n$; for instance, $\{1,2,3,1000\}$ is selected 43 times, out of 100, when $n=200$. This artificially redundant 1000th covariate acts as a ``trap'' for MPDP, which looks for the most predictive variable at each step. Moreover, this mechanism leads systematically to selecting at the first step, variable 3, and at the second step, variables 1000, for the largest sample size. Consequently, MPDP selects rarely only variables 1, 2 and 3 in any case and never for $n$ large enough whereas the performance of NOVAS increases significantly with $n$. 

Another weakness of MPDP is its propensity to retain, essentially arbitrarily, extra covariates which have no connection with the variables involved in the model; even if this trend behaves less important when $n$ increases, this still happened 34 times out of 100 for the largest sample size. The algorithm NOVAS is much less sensitive to this trap, because it allows us to select quite new models built from submodels which are not necessarily the most predictive.

\section{THEORETICAL PROPERTIES}

\ni{\sl 5.1.  Main result.}
In Theorem~1, below, we show that, with probability converging to~1 as sample size increases, the algorithm in Section~2.3 correctly determines a small, fixed number of variables on which the function $g(\cdot)=E(Y_i\mi X_i=\cdot)$ depends, even if $p$ diverges to infinity much faster than~$n$.  Moreover, the estimator $\hg$ based on these selected variables approximates $g$ with an error that, to first order, equals the error which would arise if we were told in advance the correct variables.  In this sense, $\hg$ achieves oracle performance.  

We take $\hga_{j_1,\ldots,j_\ell}$, in (2.1), to be a conventional local linear estimator in a regression on $\ell$ variables,~i.e.
$$
\hga_{j_1,\ldots,j_\ell}(x)=\bY(x)+\{\bX(x)-x\}\T\,\hSi(x)\mo\,T(x)\,,\eqno(5.1)
$$
where 
$$
\begin{array}{l}
\ds\bX(x) = {\sumi K\{(x-X_i)/h\}\,X_i\over\sumi K\{(x-X_i)/h\}}\;,\quad
\bY(x)={\sumi K\{(x-X_i)/h\}\,Y_i\over\sumi K\{(x-X_i)/h\}}\;,\\
\:\ds\hSi(x) ={\sumi\{X_i-\bX(s)\}\,\{X_i-\bX(s)\}\T\,K\{(x-X_i)/h\}\,X_i\over\sumi K\{(x-X_i)/h\}}\;,
\end{array}
$$
$K(u_1,\ldots,u_\ell)=K_1(u_1)\ldots K_1(u_\ell)$, $K_1$ is a univariate, uniformly bounded, compactly supported, symmetric probability density, and $h$ is a bandwidth.  For simplicity we use the same bandwidth for each component, although of course we could be more ambitious.

Our assumptions, (5.4), (5.5) and (5.6) are stated and discussed in Sections~5.2--5.4.  The theorem is proved in Appendix~A.  

\bs

\noindent{\bf Theorem 1.} {\em If $(5.4)$, $(5.5)$ and $(5.6)$ hold then, with probability converging to~1 as $n\rai$, the algorithm correctly concludes that $g(X_i)=E(Y_i\mi X_i)$ is a function of the first $r$ components of $X_i$ alone, and in particular the algorithm terminates at Step~$r$. } 

\bs

It is straightforward to prove from the theorem that, if the assumptions there hold, then the regression estimator based on the components to which the algorithm leads has, to first order, the same asymptotic properties as an oracle procedure based on being told in advanced that $E(Y_i\mi X_i)$ is a function of the first $r$ components of $X_i$ alone.  In particular, the asymptotic bias and variance of estimators of $g$ that are founded on the conclusion of the algorithm are first-order equivalent to their counterparts for an oracle estimator.  

\ni{\sl 5.2.  Assumptions $(5.4)$ and $(5.5)$.}
These are the main conditions for the theorem.  To state (5.4), let $f$ denote the $p$-variate density of $X$, and, given $j_1,\ldots,j_\ell$, let $\phi_{j_1,\ldots,j_\ell}(x_1,\ab\ldots,\ab x_p)$ be the $p$-variate density proportional to $f(x_1,\ldots,x_p)\,w_\ell(x_{j_1},\ab\ldots,x_{j_\ell})$, where $w_\ell$ is as in~(2.1).  Define $\psi_{j_1,\ldots,j_\ell}(x_{j_1},\ldots,x_{j_\ell})$ to be the integral of $\phi_{j_1,\ldots,j_\ell}(x_1,\ldots,x_p)$ over $x_i$ for each $i\notin\{j_1,\ldots,j_\ell\}$, and let $\ga_{j_1,\ldots,j_\ell}(x_{j_1},\ldots,x_{j_\ell})$ be the value that $E\{g(X_i)\mi X_{ij_1}\ab=x_{j_1},\ldots,X_{ij_\ell}=x_{j_\ell}\}$ would take if $X$ had density $\phi_{j_1,\ldots,j_\ell}$ rather than~$f$:
$$
\ga_{j_1,\ldots,j_\ell}(x_{j_1},\ldots,x_{j_\ell})
={\int g(x_1,\ldots,x_r)\,\phi_{j_1,\ldots,j_\ell}(x_1,\ldots,x_p)\,dx''\over
\psi_{j_1,\ldots,j_\ell}(x_{j_1},\ldots,x_{j_\ell})}\;,\eqno(5.2)
$$
where $x''$ is the $(p-\ell)$-vector that remains after $x_{j_1},\ldots,x_{j_\ell}$ have been removed from $(x_1,\ldots,x_p)$.  Define 

\begin{tabular}{cr}
\begin{minipage}{5in}
\begin{eqnarray*}
\qquad u_\ell(j_1,\ldots,j_\ell)
&=&\int\{g(x_1,\ldots,x_r)-\ga_{j_1,\ldots,j_\ell}(x_{j_1},\ldots,x_{j_\ell})\}^2\\ 
& & \qquad\qquad\qquad\quad\,\times\;
\phi_{j_1,\ldots,j_\ell}(x_1,\ldots,x_p)\,dx_1\ldots dx_p\,,\qquad\;\;(5.3)\\ 
u_0&=&E\{g(X)-Eg(X)\}^2\,. 
\end{eqnarray*}
\end{minipage}
\end{tabular}

We assume that, for a subset $\cS_\ell$ of $\IR^\ell$ that we take to be a finite union of nondegenerate compact spheres,\\ 

\begin{tabular}{cc}
\begin{minipage}{5in} (a)~Among all $\ell$-vectors $(j_1,\ldots,j_\ell)$ satisfying $1\leq j_1<\ldots<j_\ell\leq p$ and $1\leq\ell\leq r$, the choice $(1,\ldots,r)$ uniquely minimises $u_\ell(j_1,\ldots,j_\ell)$, in the sense that the minimum over all choices exceeds $u_r(1,\ldots,r)$ by at least a fixed constant~$B_3>0$, uniformly in~$n$; (b)~for some $\eta>0$, and for $1\leq\ell\leq r$, the number of distinct $\ell$-vectors $j_1,\ldots,j_\ell$, with $1\leq j_1<\ldots<j_\ell\leq p$, for which $u_\ell(j_1,\ldots,j_\ell)>n^{\eta-\{4/(\ell+4)\}}$, is of strictly smaller order than $\sqrt{q}$, and, for $1\leq\ell\leq r$, this includes all $\ell$-vectors of distinct integers chosen from $1,\ldots,r$; (c)~for each $\ell=1,2,\ldots,r+1$ there exists a constant $C_\ell>1$ such that, for all distinct $j_1,\ldots,j_\ell\in\{1,\ldots,p\}$, the joint density of $f_{j_1,\ldots,j_\ell}$ is bounded below $C_\ell$ and above $C_\ell\mo$ on $\cS_\ell$.
\end{minipage} &  (5.4)
\end{tabular}
~\\

Finally we impose basic conditions on the univariate kernel $K_1$, bandwidth~$h$ and weight function $w_\ell$ in (2.1), and on the manner in which the algorithm is terminated.  Recall that $\cS_\ell$ was introduced prior to~(5.4).\\  

\begin{tabular}{cc}
\begin{minipage}{5in}(a)~$K_1$ is a symmetric, compactly supported, H\"older continuous probability density; (b)~the bandwidth $h=h(n)$, when used to construct the $\ell$-variate regression estimator $\hga_{j_1,\ldots,j_\ell}$ at (5.1), equals a constant multiple of~$n^{-1/(\ell+4)}$; (c)~the support of the weight function $w_\ell$ equals $\cS_\ell$, and $w_\ell$ is bounded and twice differentiable there; (d)~we terminate the algorithm using the rule at (2.3), where $t=t(n)$ satisfies $n^{\eta+\{\ell/(\ell+4)\}}\leq t\leq B_5\,n$, $0<B_5<B_3$, $B_3$ is as in (5.4)(a) and $\eta$ is as in (5.4)(b). 
\end{minipage} & \mbox{(5.5)}
\end{tabular}
~\\

\ni{\sl 5.3.  Discussion of assumptions (5.4) and (5.5).}
An example where (5.4)(a) fails, and our algorithm consequently has difficulty, arises when $g(x)=x_1\ldots x_r$, the first $r$ components of $X$ are independent of one another and distributed symmetrically about zero, $w_\ell(t_1,\ldots,t_\ell)=v(t_1)\ldots v(t_\ell)$ where the nonnegative function $v$ is symmetric, and $\cS_\ell$ is a sphere centered at the origin.  Then the fitted function $\ga_{j_1,\ldots,j_\ell}$, whenever $1\leq j_1<\ldots<j_\ell\leq r$ and $1\leq\ell\leq r-1$, equals~0, and so approximating $g$ by its expected value, conditional on one or more of the first $r-1$ variables, is ineffective.  In particular, no one variable has a visible advantage over any other, and so there is no clear opportunity for choosing the correct variables.  However, this difficulty evaporates if we take the functions $w_1,\ldots,w_r$ to be sufficiently asymmetric.  This example points to the potential influence of $w_\ell$ in (2.1); for example, it can be used to counteract the  negative effects that symmetry 
has on the algorithm.  

Property (5.4)(a) implies that the choice $\ell=r$ and $j_k=k$ for $1\leq k\leq r$ uniquely minimises asymptotic mean square prediction error, but by itself (5.4)(a) does not ensure that the algorithm in Section~2.3 takes us to that particular combination of variables.  However, the latter property is guaranteed by (5.4)(b), which implies that, with high probability, the singletons $(j)$, for $1\leq j\leq r$; the doublets $(j_1,j_2)$, for $1\leq j_1<j_2\leq r$; and so on up to the $r$-tuple $(1,\ldots,r)$; are, with probability converging to~1, present among the vectors of indices treated in Steps~$1,2,\ldots,r$, respectively, in the algorithm.  Additionally, property (5.6)(d) in section~5.3, which tells us that $Y$ equals a function of the first $r$ components of $X$ alone, plus an independent error, implies that passing from these $r$ components to $r+1$ components produces, with probability converging to~1, at most a negligibly small decrease in prediction error.  In consequence, the rule (2.3) for terminating the algorithm will, with probability converging to~1, lead to a halt immediately after we have concluded, in Step~$r$, that the $r$-vector $(1,\ldots,r)$ is appropriate.  Assumption (5.5)(d), below, also helps in this regard.  

More generally, the implication from (5.6)(d) and (5.4)(a) that the choice of variable indices $(j_1,\ldots,j_\ell)=(1,\ldots,r)$ uniquely minimises $u_\ell(j_1,\ldots,j_\ell)$ allows us to investigate an oracle property of conventional type; see the theorem below.  That is, there exists a unique choice of variables that leads to best prediction of~$Y$.  Without (5.6)(d) and (5.4)(a) there could exist many different choices that produced the same asymptotic minimum mean squared prediction error.  For example, without the uniqueness part of (5.4)(a) there could exist components of the $p$-vector $X$ that were simply copies of the first $r$ components but were positioned quite differently in that vector.  On the other hand, if we were not interested in establishing such a result then we could relax (5.4) and~(5.6).  

Assumption (5.4)(b) guarantees that, although the number of variables that can be used effectively to at least partially explain $Y$ may diverge with increasing $n$, the number is not so large that extraneous variables fatally confuse the algorithm given in Section~2.3, resulting in the algorithm not correctly identifying the variable indices $(j_1,\ldots,j_\ell)=(1,\ldots,r)$ that best predict a future value of~$Y$.  

Condition (5.4)(c) asks merely that the features have the sorts of joint distributions that enable reasonable nonparametric estimation of conditional means such as $E\{g(X_i)\mi X_{ij_1},\ab\ldots,X_{ij_\ell}\}$.  To appreciate the reason for the bandwidth choices made in (5.5)(b) we note that, when computing a point estimator $\hga_{j_1,\dots,j_\ell}$ of $\ga_{j_1,\dots,j_\ell}$, the bias is of order $h^2$ (since we assumed, in (5.4)(b) and (5.4)(e), that $f_{j_1,\dots,j_\ell}$ and $g$ have two bounded derivatives), and the error about the mean is of size $(nh^\ell)\mhf$; here we used (5.4)(c).  Therefore the optimal bandwidth for point estimation of $\ga_{j_1,\dots,j_\ell}$ is of size $n^{-1/(\ell+4)}$, and that is the size assumed in~(5.5)(b).  The resulting estimator of $u_\ell$, $n\mo\,S_\ell(j_1,\dots,j_\ell)$, is in error by $O_p\{(nh^\ell)\mhf+h^2\}$.  Actually the term $(nh^\ell)\mhf$ here can be replaced by a quantity of smaller order, and the overall accuracy improved by using a smaller bandwidth, but in practice it will often be the case that the bandwidth is chosen to optimise performance for estimating $\ga_{j_1,\dots,j_\ell}$ rather than estimating $u_\ell$, and so it is appropriate to proceed as suggested above.  

\ni{\sl 5.4.  Assumption $(5.6)$.}
Condition (5.6) is standard, except perhaps for the assertion in (5.6)(d) that $g(x_1,\ldots,x_p)$ depends only on the first $r$ components $x_1,\ldots,x_r$.  However, since our algorithm is invariant under reorderings of vector components then this assumption is made without loss of generality.  It allows us to take $g$ to not depend on~$n$.\\

\begin{tabular}{cc}
\begin{minipage}{5in}
(a)~$p=p(n)$ is a function of $n$, diverging at a rate no faster than $n^{B_1}$, for some $B_1>0$, as $n$ increases; (b)~the data pairs $(X_i,Y_i)$, for $1\leq i\leq n$, are independent and identically distributed, with a common distribution that can depend on~$n$; (c)~for all choices of $j_1<\ldots<j_\ell$ from $1,\ldots,p$, each subvector $(X_{ij_1},\ldots,X_{ij_\ell})$ of $X_i$ has a well-defined probability density $f_{j_1,\ldots,j_\ell}$ (which may depend on $n$), and, for each fixed $\ell$, all second derivatives of $f_{j_1,\ldots,j_\ell}$ are bounded uniformly in distinct choices of $j_1,\ldots,j_\ell$ from $1,\ldots,p$; (d)~$Y_i=g(X_{i1},\ldots,X_{ir})+\si(X_i)\,\ep_i$, where the fixed function $g$ is uniformly bounded and has two uniformly bounded derivatives, the function $\si$ (which may depend on $n$) is uniformly bounded, and, conditional on $X_i$, the errors $\ep_i$ have a distribution depending on neither $X_i$ nor $n$, with zero mean and satisfying $E|\ep_i|^{
 B_2}<\infty$ for a sufficiently large constant $B_2>2$.
\end{minipage}
 & (5.6)
\end{tabular}

\bs\bs\ms

\cl{\bf APPENDIX~A: PROOF OF THEOREM}

\ni To simplify notation we assume throughout that the function $\si$, in (5.6)(d), is identically~1.  Let $\hga_{j_1,\ldots,j_\ell}$, $\ga_{j_1,\ldots,j_\ell}$ and $u_\ell$ be as at (5.1), (5.2) and (5.3), respectively, define $S_\ell$ as at (2.1), and let the random variable $\ep$ have the common distribution of the errors $\ep_i$ in~(5.6)(d).  Take $\eta_1$ to satisfy $0<\eta_1<\eta$, where $\eta$ is as in (5.4)(b).  Then, in view of the assumption of uniform boundedness of second derivatives in (5.6)(c) and (5.6)(d), and the assumption about the support of the density $f_{j_1,\ldots,j_\ell}$ in~(5.4)(c), 
$$
n\mo\,S_\ell(j_1,\ldots,j_\ell)=u_\ell(j_1,\ldots,j_\ell)+E\big(\ep^2\big)
+O_p\big(n^{\eta_1-\{4/(\ell+4)\}}\big)\,,\eqno(\A.1)
$$
uniformly in $1\leq j_1,\ldots,j_\ell\leq p$ and $1\leq\ell\leq r+1$.  To prove (\A.1) we need the constant $B_2$ in (5.6)(d) to be chosen sufficiently large, depending on $B_1$ in (5.6)(a) and $\eta_1$ in~(\A.1).  The proof uses Markov's inequality to bound the probability that $|n\mo\,S_\ell(j_1,\ldots,j_\ell)-\{u_\ell(j_1,\ldots,j_\ell)+E(\ep^2)\}|$ exceeds $n^{\eta_2-\{4/(\ell+4)\}}$, and observes that, since $p\leq n^{B_1}$ (see (5.6)(a)), then, for $1\leq\ell\leq r+1$, the number of vectors $(j_1,\ldots,j_\ell)$ being considered is no larger than $O(n^{(r+1)\,B_1}$).  

In Step~1 of the algorithm we take $\ell=1$ and rank values of $S_1(j)$ in order of size.  It follows from (5.4)(b) and (\A.1) that, with probability converging to~1 as $n\rai$, all the indices $j$ for which $u_1(j)\geq n^{\eta-(1/5)}$ are listed among the $\sqrt{q}$ indices for which $S_1(j)$ achieves its $\sqrt{q}$ highest values, and that this includes all the indices $1,\ldots,r$.  Similarly, in Step~2 of the algorithm we conclude with a list of ranked pairs of indices, containing all pairs chosen from among $1,\ldots,r$; and so on, until the $r$th step, when the list of selected $r$-vectors $(j_1,\ldots,j_r)$ includes $1,\ldots,r$.  It now follows from (5.4)(a) that, with probability converging to~1 as $n\rai$, the algorithm will give $(1,\ldots,r)$ the highest rank in Step~$r$, and, from (5.4)(a) and (\A.1), that with probability converging to~1 as $n\rai$ the inequality in (2.3) holds for the first time when $\ell=r$.  Therefore, with probability converging to~1 Step~$r$
  is the last step, and the $r$-tuple that is ranked most highly there, i.e.~$(1,\ldots,r)$, is the vector of component indices with which the algorithm concludes.  

\bs

\cl{\bf ACKNOWLEDGEMENT}

\ni The simulation study of this work was granted access to the HPC resources of the scientific grouping CALMIP under the allocation 2013-P1309. This grouping aims to promote the use of the new technologies in scientific computations in the researcher community of Toulouse and the french province of Midi-Pyr\'en\'ees.

\bs

\cl{\bf REFERENCES}

\noindent B\"uhlmann, P., van de Geer, S. (2011). {\em Statistics for High-Dimensional Data: Methods, Theory and Applications}. Springer.

\noindent B\"uhlmann, P., Meier, L. (2008). Discussion of ``One-step sparse estimates in nonconcave penalized likelihood models'' (authors H. Zou and R. Li). {\em Ann. Statist.}, {\bf 36}, 1534--1541.

%Borggaard, C., Thodberg, H. H. (1992). Optimal minimal neural interpretation of spectra. {\em Analytical Chemistry}, {\bf 64}, 545--551.

\noindent Bushel, P., Wolfinger, R. D., Gibson, G. (2007). Simultaneous clustering of gene expression data with clinical chemistry and pathological evaluations reveals phenotypic prototypes. {\em BMC Systems Biology} {\bf 1}(15). doi:10.1186/1752-0509-1-15.

\noindent Cand\`es, E., Tao, T. (2007). The Dantzig selector: statistical estimation when $p$ is much larger than $n$. {\em Ann. Statist.}, {\bf 35}, 2313--2351.

\noindent Dejean, S., Gonzalez, I., Le Cao, K.-A., Monget, P., Coquery, J. (2011). mixOmics: Omics Data Integration Project. R package version 3.0.
  http://CRAN.R-project.org/ package=mixOmics.

\noindent Efron, B., Hastie, T., Johnstone, I., Tibshirani, R. (2004). Least angle regression. {\em Ann. Statist.}, {\bf 32}, 407--499.

\noindent Fan, J., Gijbels, I, (1996). {\em Local Polynomial Modeling and its Applications}. Chapman and Hall, London.

\noindent Fan, J., Li, R. (2001). Variable selection via nonconcave penalized likelihood and its oracle properties. {\em J. Am. Statist. Ass.}, {\bf 96}, 1348--1360.

\noindent Fan, J., Lv, J. (2010). A selective overview of variable selection in high dimensional feature space (invited review article). {\em Statistica Sinica}, {\bf 20}, 101--148.

\noindent Ferraty, F., Vieu, P. (2002). The functional nonparametric model and application to spectrometric data. {\em Computational Statistics}, {\bf 17}, 545--564.

\noindent Ferraty, F., Hall, P., Vieu, P. (2010). Most-predictive design points for functional data predictors. {\em Biometrika}, {\bf 97}, 807--824.

%Ferraty, F., Vieu, P. (2006). {\em Nonparametric functional data analysis}. Springer, New York.

\noindent Friedman, J., Hastie, T., H\"ofling, H. and Tibshirani, R. (2007). Pathwise coordinate optimization. {\em Ann. Appl. Statist.}, {\bf 1}, 302--332.

\noindent Fu, W. (1998). Penalized regressions: the Bridge versus the lasso. {\em Journal of  Computational and Graphical Statistics}, {\bf 7}, 397--416.

\noindent Geladi, P., Kowalski, B.R. (1986). Partial least squares regression: A Tutorial. {\em Analytica Chimica Acta}  {\bf 185}, 1--17.

\noindent Hastie, T., Tibshirani, R., Friedman, J. (2009). {\em The Elements of Statistical Learning: Data Mining, Inference, and Prediction} (2nd edition). Springer, New York.

%Hayfield, T., Racine, J.S. (2008). Nonparametric Econometrics: The np Package. {\em J. Statist. Softw.} 27(5). URL http://www.jstatsoft.org/v27/i05/.

\noindent Huang, J., Horowitz, J. L., Wei, F. (2010). Variable selection via nonparametric additive models. {\em Ann. Statist.}, {\bf 38}, 2282--2313.

\noindent L\^e Cao, K. A., Rossouw D., Robert-Grani\'e, C., Besse, P. (2008). A sparse PLS for variable selection when integrating Omics data. {\em Statist. Appl. Genet. Mol. Biol.} {\bf 7}, article 35.

\noindent Martens, H., Naes, T. (1989). {\em Multivariate calibration}. John Wiley \& Sons Ltd.

\noindent Meier, L., van de Geer, S., B\"uhlmann, P. (2009). High-dimensional additive modeling. {\em Ann. Statist.}, {\bf 37}, 3779--3821.

\noindent Meinshausen, N. (2007). Relaxed lasso. {\em Computational Statistics and Data Analysis}, {\bf 52}, 374--393.

\noindent R Development Core Team (2011). R: A language and environment for statistical computing. R Foundation for Statistical Computing, Vienna, Austria. ISBN 3-900051-07-0, URL http://www.R-project.org/.

\noindent Ravikumar, P., Lafferty, J., Liu, H., Wasserman, L. (2009). Sparse additive models. {\em J. R. Statist. Soc.} B, {\bf 71}, 1009--1030.

\noindent Revolution Analytics (2011). doSNOW: Foreach parallel adaptor for the snow package. R package version 1.0.5. (Available from http://CRAN.R-project.org/package =doSNOW.)

\noindent Tibshirani, R. (1996). Regression analysis and selection via the lasso. {\em J. R. Statist. Soc.} B, {\bf 58}, 267--288.

\noindent Wold, H. (1966). Estimation of principal components and related models by iterative least squares. In: Krishnaiah, P.R. (editors). {\em Multivariate Analysis}. Academic Press, N.Y., 391--420.

\noindent Yuan, M., Lin, Y. (2006). Model selection and estimation in regression with grouped variables. {\em J. R. Statist. Soc.} B, {\bf 68}, 49--67.

\noindent Zou, H. (2006). The adaptive lasso and its oracle properties. {\em J. Am. Statist. Ass.}, {\bf 101}, 1418--1429.

\noindent Zou, H., Hastie, T. (2005). Regularization and variable selection via the Elastic Net. {\em J. R. Statist. Soc.} B, {\bf 67}, 301--320.

\end{document}